\documentclass[10pt]{article}
\usepackage{graphicx}
\usepackage{amsmath}
\usepackage{amssymb}
\usepackage{caption2}
\begin{document}
\begin{center}
{\large {\bf \sc{ Comment on "Hybridized Tetraquarks" }}} \\[2mm]
Zhi-Gang Wang \footnote{E-mail: zgwang@aliyun.com.  }     \\
 Department of Physics, North China Electric Power University, Baoding 071003, P. R. China
\end{center}

\begin{abstract}
In this comment, I illustrate that the formula $\Gamma=A \sqrt{\delta}$  suggested in  arXiv:1603.07667 comes from a kinematical factor, and  has no relation to the existence or non-existence of the $X(5568)$.
\end{abstract}

In the two-body strong decays $A\to BC$, the partial decay width can be written as
\begin{eqnarray}
\Gamma&=&\frac{1}{2j_A+1}\frac{p}{8\pi m_A^2} |T|^2 \, ,
\end{eqnarray}
where
\begin{eqnarray}
p&=&\frac{\sqrt{\left[m_A^2-(m_{B}+m_{C})^2 \right]\left[m_A^2-(m_{B}-m_{C})^2 \right]} }{2m_A} \, ,
\end{eqnarray}
is the three momentum of the final mesons in the  center of mass
of the initial  meson $A$, the $j_A$ is the spin of the initial  meson $A$, the $T$ is the scattering amplitude.
If there exists a relative angular momentum $L$ between the final mesons $B$ and $C$, then
\begin{eqnarray}
\Gamma &\sim& p^{2L+1}\, .
\end{eqnarray}
Now we set $L=0$ for simplicity.

In the case $m_{B}\approx m_{C}$,
\begin{eqnarray}
p&\approx &\frac{\sqrt{m_A+(m_{B}+m_{C}) } \sqrt{m_A-(m_{B}+m_{C}) } }{2} =\frac{\sqrt{2\left(m_B+m_C\right)+\delta}}{2}\sqrt{\delta}\nonumber\\
&\approx& \frac{\sqrt{2(m_B+m_C)}}{2}\sqrt{\delta}\, ,
\end{eqnarray}
where $\delta=m_A-(m_{B}+m_{C})\ll m_B+m_C$.
The three momentum $p$ in the decays
\begin{eqnarray}
X(3872) &\to& \bar{D}^0\, D^{*0}\, , \nonumber\\
Z_c^+(3900) &\to& \bar{D}^0\, D^{*+}\, , \nonumber\\
Z_c^+(4025) &\to &\bar{D}^{*0}\, D^{*+}\, , \nonumber\\
Z_b^+(10610) &\to& \bar{B}^0 \, B^{*+}\, ,\nonumber \\
Z_b^+(10650) &\to &\bar{B}^{*0}\, B^{*+}\, ,
\end{eqnarray}
can be well approximated by Eq.(4) \cite{PDG}. The  partial decay widths
can be well  fitted into the following form,
\begin{eqnarray}
\Gamma&=&A\sqrt{\delta}\, ,
\end{eqnarray}
 where $A$ is a fitted parameter \cite{EPP-1603}. The formula in Eq.(6) comes from a kinematical factor,  the hybridization mechanism proposed in Ref.\cite{EPP-1603} is unnecessary. In fact, those partial decay widths have  not been measured yet, even the total widths have not been precisely  measured, some decays have not   been observed yet \cite{PDG}. We can only say that the partial decay widths are of the form $\Gamma=A\sqrt{\delta}$, as the input parameters  are not robust. 

In the case $m_{B}\gg m_{C}$,
\begin{eqnarray}
p&= &\frac{\sqrt{m_A+(m_{B}+m_{C}) } \sqrt{m_A+(m_{B}-m_{C}) } \sqrt{m_A-(m_{B}+m_{C}) } \sqrt{m_A-(m_{B}-m_{C}) }}{2m_A}  \nonumber\\
&\approx& \frac{m_A+m_B}{2m_A}\sqrt{\delta}\,\sqrt{\delta+2m_C} \, ,
\end{eqnarray}
where $\delta=m_A-(m_{B}+m_{C})\ll m_B+m_C$. The three momentum $p$ in the decays
\begin{eqnarray}
Y(4140) &\to& J/\psi\, \phi\, , \nonumber\\
Z(4430) &\to &\eta_c^{\prime}\,\rho \, , \\
X(5560) &\to & B_s \, \pi \, ,
\end{eqnarray}
can be well approximated by Eq.(7), not by Eq.(4). The decay $Z(4430) \to \eta_c^{\prime}\,\rho$ has not been observed yet, and the partial decay width of the process  $Y(4140) \to J/\psi\, \phi$ has not been measured \cite{PDG}. The  partial decay widths
can be  written  into the  form,
\begin{eqnarray}
\Gamma&=&A\sqrt{\delta}\,\sqrt{\delta+A^\prime} \, ,
\end{eqnarray}
not the form in Eq.(6), where the $A$ and $A^{\prime}$ are fitted parameters. In the case $\delta \ll A^{\prime}$, Eq.(10) is reduced to Eq.(6). Here I assume the spin-parity $J^P$ of the $Y(4140)$ is $0^+$ or $2^+$. Again the formula in Eq.(10) comes from a kinematical factor,  the hybridization mechanism proposed in Ref.\cite{EPP-1603} is unnecessary.  In Ref.\cite{EPP-1603}, the widths of the $Y(4140)$, $Z(4430)$ and $X(5568)$ cannot be well approximated by Eq.(6).

In the chiral limit $m_C=m_{\pi}\to 0$,
\begin{eqnarray}
p&\approx& \frac{m_A+m_B}{2m_A}\, \delta\, .
\end{eqnarray}
The three momentum $p$ in the decays
\begin{eqnarray}
Z_c(3900) &\to& J/\psi \, \pi\, , \nonumber\\
Z(4430) &\to& \psi^{\prime} \, \pi\, ,\\
X(5560) &\to & B_s \, \pi \, ,
\end{eqnarray}
can be approximated by Eq.(11). The  partial decay widths
can be written into the following form,
\begin{eqnarray}
\Gamma&=&A\,\delta \, ,
\end{eqnarray}
not the form in Eq.(6), where the $A$ is fitted parameter. Again the formula in Eq.(14) comes from a kinematical factor.
  
  In summary, the $\delta$ dependence of the partial decay widths comes from the kinematical factors,  the hybridization mechanism proposed in Ref.\cite{EPP-1603} is unnecessary, and has no relation to the existence or non-existence of the $X(5568)$ \cite{X5568-exp}.

\section*{Acknowledgements}
This  work is supported by National Natural Science Foundation,
Grant Number 11375063, and Natural Science Foundation of Hebei province, Grant Number A2014502017.

\end{document}